\documentclass[runningheads]{llncs}
\usepackage[T1]{fontenc}
\usepackage{graphicx}
\usepackage{amsmath}
\usepackage{cite} 
\usepackage{amssymb}
\usepackage{hyperref}
\usepackage{booktabs}
\usepackage{adjustbox}
\usepackage[table,xcdraw]{xcolor}
\usepackage{subcaption}
\usepackage{multirow}
\usepackage{comment}
\usepackage{booktabs} 
\usepackage{siunitx}  

\begin{document}
\title{CT2Rep: Automated Radiology Report Generation for 3D Medical Imaging}
%
%

\author{Ibrahim Ethem Hamamci\inst{1} \and
Sezgin Er\inst{2} \and
Bjoern Menze\inst{1}}  


\authorrunning{Hamamci et al.}

\institute{\begin{tabular}{c}
{$^{1}$University of Zurich \quad $^{2}$Istanbul Medipol University}
\end{tabular}\\
\email{\{ibrahim.hamamci@uzh.ch\}}}

\maketitle              
\begin{abstract}
Medical imaging plays a crucial role in diagnosis, with radiology reports serving as vital documentation. Automating report generation has emerged as a critical need to alleviate the workload of radiologists. While machine learning has facilitated report generation for 2D medical imaging, extending this to 3D has been unexplored due to computational complexity and data scarcity. We introduce the first method to generate radiology reports for 3D medical imaging, specifically targeting chest CT volumes. Given the absence of comparable methods, we establish a baseline using an advanced 3D vision encoder in medical imaging to demonstrate our method's effectiveness, which leverages a novel auto-regressive causal transformer. Furthermore, recognizing the benefits of leveraging information from previous visits, we augment CT2Rep with a cross-attention-based multi-modal fusion module and hierarchical memory, enabling the incorporation of longitudinal multimodal data. Access our code at \url{https://github.com/ibrahimethemhamamci/CT2Rep}.


\end{abstract}

\section{Introduction}
\label{section:introduction}

The integration of machine learning into radiology, driven by numerous public datasets \cite{wang2017chestx, irvin2019chexpert, hamamci2023dentex}, has significantly enhanced disease classification and segmentation \cite{Yuksel2021, pati2023gandlf, 10.1007/978-3-031-43987-2_38, draelos2021machine}. Furthermore, recent advancements have enabled the development of many methods for generating radiology reports for 2D medical imaging \cite{chen2020generating,thirunavukarasu2023large, li2023dynamic, jing2017automatic, wang2022cross} utilizing public datasets \cite{johnson2019mimic, nguyen2022vindr}. However, this progress in report generation has not yet extended to 3D medical imaging, due to computational complexities \cite{gao2022get3d} and the lack of datasets paired with radiology reports \cite{chen2022recent}.

3D medical imaging, such as computed tomography (CT) and magnetic resonance imaging, provides a more detailed perspective on the patient's condition compared to 2D imaging \cite{muller2002computed}. Consequently, manual report generation, essential for conveying diagnostic findings, becomes more time-consuming and error-prone, highlighting the need for automation. One of the challenges in developing such a framework lies in the scarcity of 3D medical imaging datasets paired with reports \cite{li2023systematic}. Moreover, the nature of 3D images involves volumetric data, which necessitates more sophisticated algorithms for interpreting the additional dimension. This complexity presents unique obstacles to generating descriptive and clinically relevant reports that effectively capture the details of 3D images.

Recognizing this gap, our work introduces CT2Rep, the first approach to automated radiology report generation for 3D medical imaging, specifically targeting chest CT volumes. CT2Rep leverages a novel 3D auto-regressive causal vision feature extractor, optimized for processing volumetric data. We also incorporate relational memory to utilize information from previous report generations, employing memory-driven conditional layer normalization to integrate this data into our framework. To train our framework, we utilize the CT-RATE dataset \cite{hamamci2024foundation}, which consists of 25,692 non-contrast chest CT volumes, expanded to 50,188 through various reconstructions, from 21,304 unique patients, along with corresponding radiology reports. CT2Rep's uniqueness, the first of its kind in 3D medical imaging, means that no directly comparable methods exist. Nonetheless, to demonstrate the effectiveness of our framework, we reasonably designed a baseline using a state-of-the-art vision encoder used for 3D chest CT volume interpretation, CT-Net \cite{draelos2021machine}, for report generation. CT2Rep outperforms this well-designed baseline method, showcasing the efficacy of our novel approach.

Radiologists typically assess a 3D chest CT volume alongside previous volumes and reports for the same patient, as multiple visits are common in clinical practice. Longitudinal volumes and their reports contain valuable information, and leveraging this multimodal data can potentially enhance report generation. Hence, we extended CT2Rep by incorporating a cross-attention-based multi-modal fusion module coupled with a hierarchical memory-driven decoder. This extension not only addresses computational challenges associated with 3D image analysis but also facilitates the inclusion of longitudinal multimodal patient data, enriching the context and accuracy of generated reports. We evaluated this extended version, named CT2RepLong, through a comprehensive ablation study to underscore the importance of historical imaging and reports in informing current diagnostic interpretations. Our contributions can be summarized as:

\begin{itemize}
\item We propose CT2Rep, the first radiology report generation framework for 3D medical imaging, employing a novel auto-regressive causal transformer.
\item As CT2Rep is the first of its kind and no comparable methods exist, we have designed a baseline employing the cutting-edge 3D vision encoder used in chest CT classification to benchmark our method and prove its effectiveness.
\item We augment CT2Rep with a cross-attention-based multi-modal fusion module and a hierarchical memory-driven decoder to leverage commonly available longitudinal data, backed by a comprehensive ablation study showcasing the efficacy of incorporating longitudinal data for report generation.
\item We make our trained models and source codes publicly available to facilitate out-of-the-box report generation for 3D chest CT volumes.
\end{itemize}

\section{Methods}
\label{section:methods}
Although 3D medical imaging, such as 3D chest CT volumes, offers more comprehensive information than its 2D counterparts like chest X-rays, there are currently no solutions for generating radiology reports for 3D imaging due to data scarcity and computational complexity. To address this gap, we developed a 3D sequence-to-sequence generation model, detailed in Sec. \ref{CT2Rep}, utilizing the data outlined in Sec. \ref{data}. Additionally, we enhanced our method to incorporate longitudinal multimodal data from previous visits, as described in Sec. \ref{CT2RepLong}.

\subsection{The Proposed Method}
\label{CT2Rep}
\begin{figure}

    \centering
    \includegraphics[width=1\linewidth]{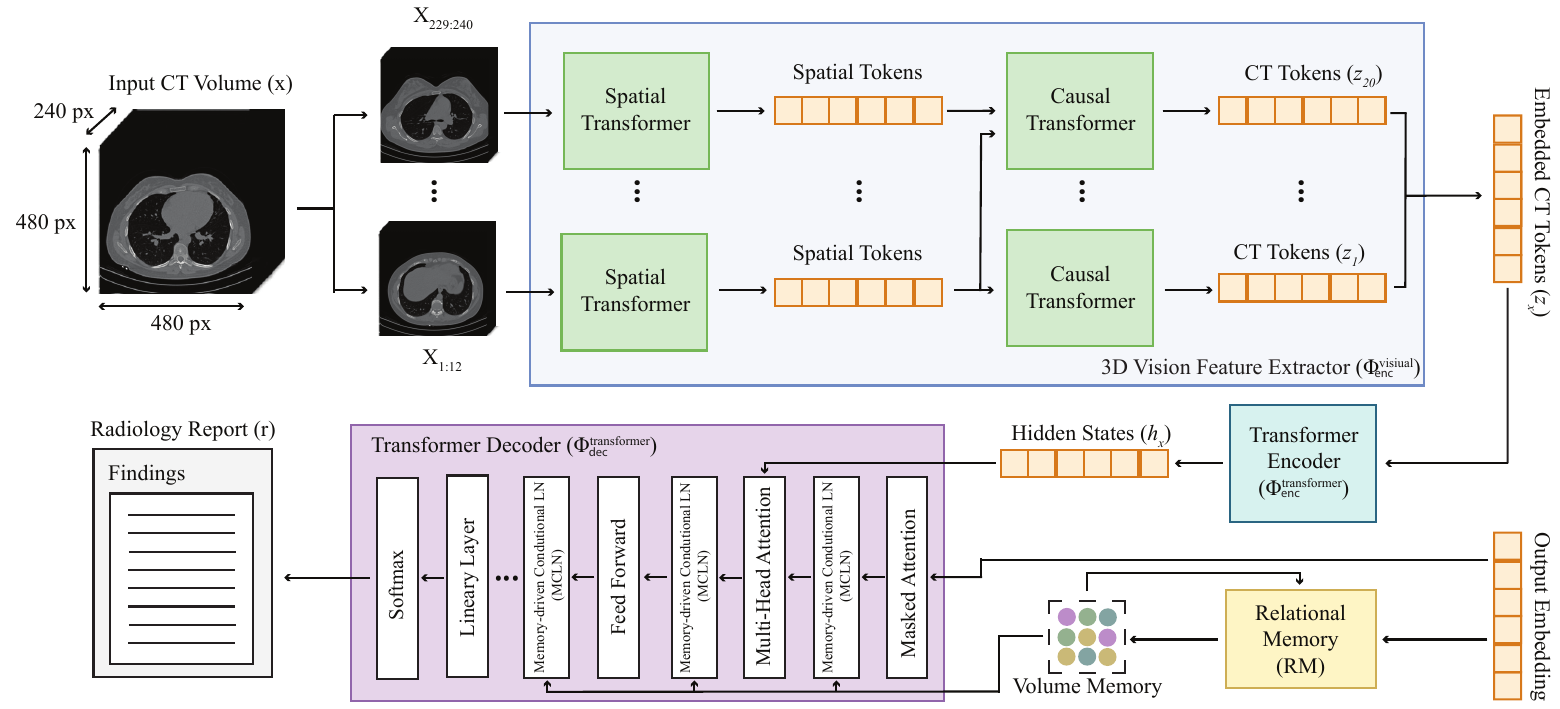}
    \caption{CT2Rep features a novel auto-regressive causal transformer for 3D vision feature extraction, complemented by RM and MCLN-enhanced transformer-based encoder and decoder network for clinically accurate report generation.}
    \label{ct2rep}

\end{figure}

The model ($\Phi_\text{CT2Rep}$) accepts input 3D volumes $x\in{\mathbb{R}^{(240)\times 480\times 480}}$ as a sequence of CT patches $x = \{ x_1, x_2, x_3,..., x_N\}, x_n\in\mathbb{R}^{(12)\times 24 \times 24} $ to predict a target sequence $r_{out} = \{r_1^\text{out}, r_2^\text{out},..., r_T^\text{out}\}, r_t^\text{out} \in \mathbb{V}$. Here, $N$ represents CT feature count, $T$ the token count, and $\mathbb{V}$ the possible token vocabulary. CT2Rep, depicted in Fig. \ref{ct2rep}, consists of three key components, each elaborated below.

\noindent\textbf{3D vision feature extractor.} A key component and main contribution of our framework, this network ($\Phi^\text{visual}_\text{enc}$) facilitates the extraction of embedded CT tokens from 3D chest CT volumes by segmenting the data into distinct patches and transforming them into a lower-dimensional latent space, inspired by \cite{arnab2021vivit}. These tokens capture essential information, facilitating subsequent analysis. 

The network takes a 3D CT volume ($x$) and produces embedded CT tokens $z_{x}\in{\mathbb{R}^{20\times 20\times 20 \times 512}}$, by initially extracting \( (12) \times 24 \times 24\) non-overlapping patches from $x$. Each patch is then mapped to a \(D\)-dimensional space, with \(D\) set to $512$. The patches are then reshaped and transformed linearly to \( B \times T \times \frac{H}{p_1} \times \frac{W}{p_2} \times D \), following a previous work \cite{hamamci2023generatect}. Here, \( p_t \) denotes the temporal patch size, \( T \) represents the number of temporal patches, \( B \) is the batch size, \( H \) and \( W \) are the height and width of the slices, respectively, and \( p_1 \) and \( p_2 \) represent the spatial patch sizes. After patch embedding, the resulting tensor size is \( B \times (T) \times \frac{H}{p_1} \times \frac{W}{p_2} \times D \). This tensor is then processed by two transformer networks consecutively. First, the spatial transformer operates on a reshaped tensor of size \( (B \cdot (T)) \times (\frac{H}{p_1} \cdot \frac{W}{p_2}) \times D \), yielding a tensor of the same dimensions. Subsequently, the causal transformer processes this output reshaped to \( (\frac{H}{p_1} \cdot \frac{W}{p_2}) \times (B \cdot (T)) \times D \), and produces an output maintaining these dimensions. This method ensures that both spatial and latent dimensions are preserved after each layer, thereby retaining 3D volumetric information throughout the network's processing stages. The overall 3D chest CT volume feature extraction process, formally defined as $z_{x} = \Phi^\text{visual}_\text{enc}(x)$, ensures 3D volumetric information is preserved, facilitating the effective construction of sequence-to-sequence models for report generation.

\noindent\textbf{Transformer encoder.} We employ a conventional transformer ($\Phi^\text{transformer}_\text{enc}$) to encode CT features extracted by $\Phi^\text{visual}_\text{enc}$. This network processes these features to produce encoded hidden states via an attention mechanism, crucial for capturing feature interdependencies. The encoded hidden states are represented as:
\begin{equation*} h_x = \{h_1, h_2, \ldots, h_N\} = \Phi^\text{transformer}_\text{enc} (z_x) = \Phi^\text{transformer}_\text{enc} (z_1, z_2, \ldots, z_N),\end{equation*}
where each $h_n \in \mathbb{R}^{512}$ represents the encoded state of a patch, with $N$ being the total patch count. The attention mechanism in the transformer is defined as $\text{Attention}(Q, K, V) = \text{softmax}\left(\frac{QK^T}{\sqrt{d_k}}\right)V$, where $Q$, $K$, and $V$ stand for the query, key, and value matrices, respectively, and $d_k$ is the key's dimensionality.

\noindent\textbf{Transformer decoder.} We adapt a traditional transformer network as a decoder ($\Phi^\text{transformer}_\text{dec}$), with two notable enhancements. First, we integrate relational memory (RM) \cite{chen2020generating}, entailing the utilization of a matrix to encapsulate and propagate pattern information across generation steps. Each row within this matrix stores specific pattern details, which are iteratively refined through updates incorporating outputs from preceding steps. The updating mechanism involves employing the matrix from the previous step as a query and concatenating it with the prior output to serve as the key and value for the transformer's multi-head attention module. Mathematically, this process is achieved through multi-head attention, where \(Q = M_{t-1} \cdot W_q\), \(K = [M_{t-1}; y_{t-1}]\cdot W_k\), and \(V = [M_{t-1}; y_{t-1}] \cdot W_v\). Here, \(y_{t-1}\) denotes the embedding of the previous step's output, while \(W_q\), \(W_k\), and \(W_v\) represent the trainable weights for query, key, and value transformations, respectively. Thus, the model effectively learns conserved report patterns, such as \emph{"Trachea, both main bronchi are open."}, within similar CT volumes. Second, we employ a memory-driven conditional layer normalization (MCLN) \cite{lample2019large}, integrating RM directly into the decoder's scaling (\(\gamma\)) and shifting (\(\beta\)) parameters. This makes the model more contextually aware and adept at generating accurate text outputs. The decoding process is defined as:
\begin{equation*}
r^\text{out}_T = \Phi^\text{transformer}_\text{dec} (h_1, h_2, \ldots, h_N, MCLN(RM(r^\text{out}_{1}, r^\text{out}_{2}, \ldots, r^\text{out}_{T-2},r^\text{out}_{T-1}))).
\end{equation*}
\noindent\textbf{Inference.}
After training, CT2Rep ($\Phi_\text{CT2Rep}$) is able to generate a radiology report ($r_\text{out}$) for a given 3D chest CT volume ($x$), formally defined as follows:
\begin{equation*}
r_\text{out} = \Phi_\text{CT2Rep}(x) =  \Phi^\text{transformer}_\text{dec} ( \Phi^\text{transformer}_\text{enc}( \Phi^\text{visual}_\text{enc}(x))).
\end{equation*}

\subsection{Longitudinal Data Utilization}\label{CT2RepLong}
\begin{figure}
    \centering
    \includegraphics[width=1\linewidth]{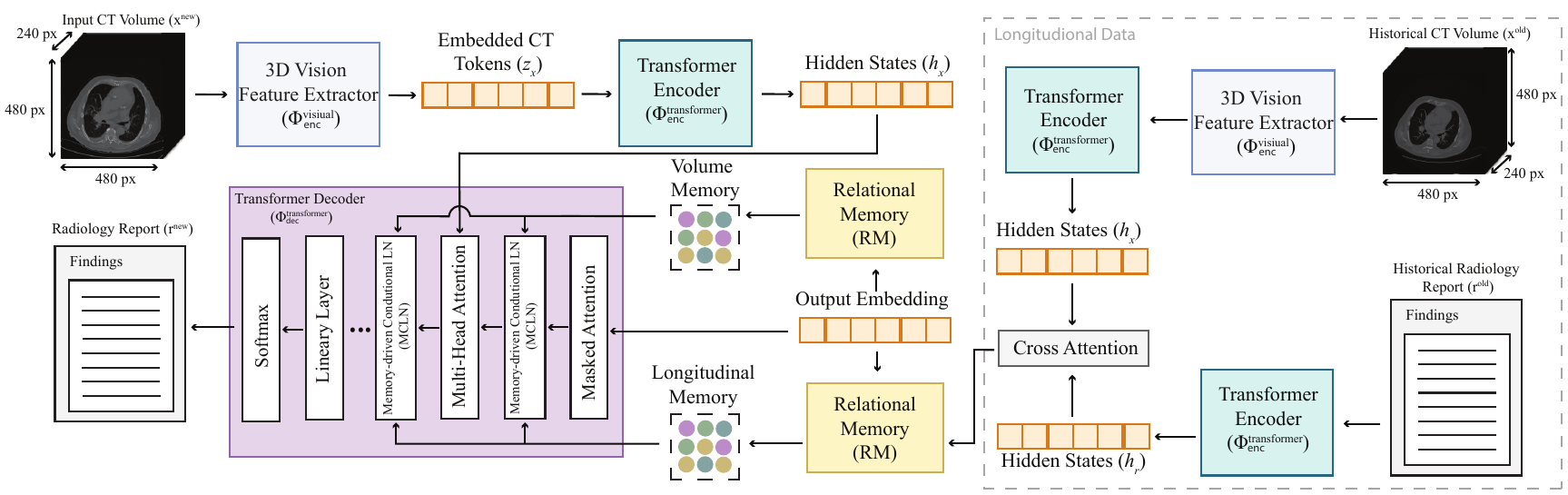}
    \caption{CT2RepLong enhances CT2Rep with a cross-attention multi-modal fusion module and longitudinal memory for effective historical data integration.}
    \label{ct2replong}
    \end{figure}

To utilize multimodal data from previous visits, we augmented CT2Rep with a cross-attention-based fusion module \cite{zhu2023utilizing} that allows to predict an outcome sequence $r_{out}^{new} = \{r_1, r_2,..., r_T\}, r_t \in \mathbb{V}$, for a given new 3D chest CT volume ($x^{new}$) by integrating representations from the previous CT volume ($x^{old}$) and its corresponding previous report ($r_{in}^{old}$). The fusion process is facilitated by computing cross-attention between previous volume and report representations by $R^*=\text{softmax}\left(\frac{q(H_{RP})k(H_{IP})^\top}{\sqrt{d_k}}\right)$ and $I^*=\text{softmax}\left(\frac{q(H_{IP})k(H_{RP})^\top}{\sqrt{d_k}}\right)$, where $H_{IP}$ and $H_{RP}$ are the attended features for longitudinal volumes and reports, respectively. These features are concatenated to create a comprehensive multimodal longitudinal representation $H_L$. This integrated approach significantly enhances the performance of the longitudinal framework, $\Phi_\text{CT2RepLong}$, by leveraging both spatial and semantic information from previous visits, as detailed in Fig. \ref{ct2replong}.

\subsubsection{Multimodal transformer decoder.}
The decoder of $\Phi_\text{CT2RepLong}$ closely follows that of $\Phi_\text{CT2Rep}$. However, $\Phi_\text{CT2RepLong}$ employs two more cross-attention mechanisms, together defined as $\Phi^\text{attn}_\text{long}$, to analyze the relationships between previous reports ($r^\text{old}$) and volumes ($x^\text{old}$), and vice versa. The outputs are concatenated, and then RM is applied to the new report as per the process described in Sec. \ref{CT2Rep}. Subsequently, another cross-attention, ($\Phi^\text{attn}_\text{mem}$), is used between the RM and the cross-attention outputs from previous volumes and reports. The resulting cross-attention outputs are then utilized in MCLN, formalized as:
\begin{equation*}
\resizebox{\linewidth}{!}{$
r^\text{out}_T = \Phi^\text{transformer}_\text{dec} (h_1,..., h_N, MCLN(\Phi^\text{attn}_\text{mem}(RM(r^\text{out}_{1},..., r^\text{out}_{T-1}), \Phi^\text{attn}_\text{long}(r^\text{old}, x^\text{old})))).
$}
\end{equation*}
\subsubsection{Inference.}
After training, $\Phi_\text{CT2RepLong}$ can generate a report ($r_\text{out}^\text{new}$) for a given new volume ($x^\text{new}$), alongside a previous volume and its corresponding report:
\begin{equation*}
\begin{aligned}
r_\text{out}^\text{new} &= \Phi_\text{CT2RepLong}(x^\text{new}, x^\text{old}, r^\text{old}) \\
&= \Phi^\text{transformer}_\text{dec} ( \Phi^\text{transformer}_\text{enc}( \Phi^\text{visual}_\text{enc}(x^\text{new})), \Phi^\text{attn}_\text{long}(r^\text{old}, x^\text{old})).
\end{aligned}
\end{equation*}

\subsection{Dataset Preparation.}\label{data} We utilize 3D chest CT volumes along with corresponding radiology reports from the publicly available CT-RATE dataset \cite{hamamci2024foundation}. For the development of CT2Rep, we employ all volumes and reports from the initial release of CT-RATE. Our dataset comprises 25,701 non-contrast 3D chest CT volumes from 21,314 unique patients, which expands to 49,138 volumes after applying multiple reconstructions tailored to different window settings \cite{willemink2019evolution}. Each volume features a resolution of $512 \times 512$ pixels in the axial plane, with slice counts ranging from 100 to 600. The radiology reports associated with each volume are segmented into four sections: clinical information, technique, findings, and impression; however, only the findings section is utilized for report generation training. The same radiology report is used for each reconstructed volume of a single CT volume. The dataset is divided into a training set of 20,000 patients and a validation set of 1,314 patients, ensuring no overlap. CT volumes were converted to Hounsfield Units (HU) using slope and intercept values from the metadata and clipped to $[-1000~\text{HU}, +200~\text{HU}]$ to represent the practical diagnostic limits of the HU scale \cite{denotter2019hounsfield}. Each volume was subsequently resized to achieve uniform spacing of 0.75 mm on the x and y axes and 1.5 mm on the z-axis. The volumes were either center-cropped or padded to achieve a consistent resolution of $(240) \times 480 \times 480$. 

\noindent\textbf{Creating the longitudinal dataset.} We targeted patients with more than two visits, yielding 6,766 and 429 3D chest CT volumes from 2,638 and 169 unique patients for the training and validation sets, respectively. After applying various reconstructions, these volumes increased to 13,354 for training and 849 for validation. We chronologically ordered the volumes for each patient using the \emph{StudyTime} metadata attribute and paired every two possible longitudinal volumes for a patient, resulting in 28,441 training and 1,689 validation pairs.

\section{Experiments and Results}
\label{section:results}
\definecolor{mycolor}{rgb}{0.90, 0.90, 0.90} 

\begin{table}[t]
  \caption{Quantitative evaluation showcases the effectiveness of our CT2Rep model in comparison to a well-designed baseline and highlights how our enhanced method, CT2RepLong, leverages longitudinal data to improve performance.}
  \centering
  \setlength{\tabcolsep}{2.5pt} 
  \renewcommand{\arraystretch}{1.05} 
  \begin{tabular}{
    @{}
    l
    c S[table-format=4.1]
    S[table-format=1.4]
    S[table-format=3.1]
    S[table-format=2.1]
    c S[table-format=1.4]
    S[table-format=3.1]
    S[table-format=2.1]
    @{}
  }
    \toprule
    & \multicolumn{6}{c}{\textbf{NLG Metrics}} & \multicolumn{3}{c}{\textbf{CE Metrics}} \\
    \cmidrule(lr){2-7} \cmidrule(lr){8-10}
    \textbf{Method} & \textbf{BL-1} & \textbf{BL-2} & \textbf{BL-3} & \textbf{BL-4} & \textbf{M} & \textbf{R\(_{\textbf{L}}\)} & \textbf{P} & \textbf{R} & \textbf{F1} \\
    \midrule
    Base w/ CT-Net & {0.443} & {0.399} & {0.375} & {0.354} & {0.286} & {0.442} &  {0.513} & {0.531} & {0.456} \\ 
    \rowcolor{mycolor} 
    CT2Rep (Ours) & \textbf{0.460} & \textbf{0.415} & \textbf{0.390} & \textbf{0.369} & \textbf{0.295} & \textbf{0.459} & \textbf{0.749} & \textbf{0.548} & \textbf{0.534} \\
    \midrule
    \multicolumn{10}{l}{\textbf{\textit{methods below utilize longitudinal data}}} \\
    Baseline & {0.372} & {0.317} & {0.282} & {0.251} & {0.238} & \textbf{0.353} & {0.666} & {0.465} & {0.525} \\ 
    $+$ report & {0.330} & {0.284} & {0.260} & {0.241} & {0.213} & {0.313} & {0.623} & {0.410} & {0.524} \\ 
    $+$ volume & {0.305} & {0.261} & {0.238} & {0.220} & {0.204} & {0.291} & {0.662} & {0.434} & {0.530} \\ 
    $+$ report $+$ volume & {0.365} & {0.319} & {0.292} & {0.271} & {0.239} & {0.351} & {0.658} & {0.410} & {0.533} \\ 
    \rowcolor{mycolor} 
    CT2RepLong (Ours) & \textbf{0.374} & \textbf{0.327} & \textbf{0.304} & \textbf{0.401} & \textbf{0.285} & {0.263} & \textbf{0.727} & \textbf{0.511} & \textbf{0.536} \\ 
    \bottomrule
  \end{tabular}

  \label{table}

\end{table}

To evaluate model efficacy in generating radiology reports, we employed natural language generation (NLG) and clinical efficacy (CE) metrics. NLG metrics include BLEU (BL) \cite{papineni2002bleu}, METEOR (M) \cite{lavie2009meteor}, and ROUGE-L (R\(_{\textbf{L}}\)) \cite{lin2004rouge}, assessing word overlap, synonym use and word order, and sequence matching, respectively. For CE metrics, we fine-tuned the CXR-Bert model \cite{boecking2022making} for multi-label classification of reports on 18 abnormalities, as detailed in the supplementary material. We then predicted the abnormality labels of both ground-truth data and generated reports and computed classification scores, including precision (P), recall (R), and F1 score, to measure the clinical accuracy of the generated reports.

\subsection{Comparison with the Baseline Method} \label{base}
Given the absence of directly comparable methods, further highlighting our method's novelty, we established a benchmark for radiology report generation by implementing a state-of-the-art vision encoder, CT-Net \cite{ctnet}, used in 3D medical imaging. CT-Net is the first and only model developed for classifying 3D chest CT volumes. Its architecture comprises a ResNet-18 feature extractor \cite{resnet}, augmented by 3D convolutional blocks designed to streamline ResNet features, followed by final classification layers. In our approach, we harness the feature extraction capabilities of CT-Net, using these features as inputs for our 3D volume transformer, establishing it as the baseline for our study. Table \ref{table} demonstrates that our CT2Rep significantly outperforms this baseline, thanks to our novel auto-regressive causal transformer used as the 3D vision feature extractor.

\begin{figure}
    \centering
    \includegraphics[width=1\linewidth]{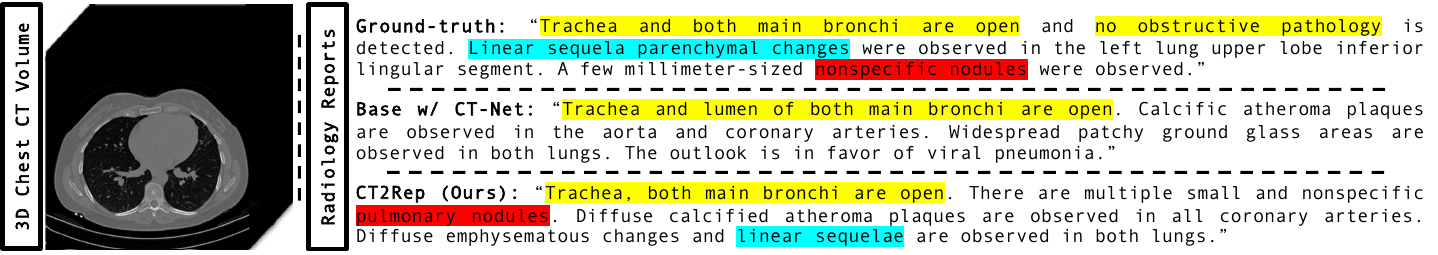}
    \caption{Comparison of ground-truth with reports generated by a CT-Net-based baseline and CT2Rep, highlighting CT2Rep's medical precision with color codes.}
    \label{ct2repcase}
\end{figure}

\noindent\textbf{Case study.} We assessed our model's performance through a qualitative analysis on a randomly chosen case from our test set, comparing generated reports to the ground truth. Figure \ref{ct2repcase} illustrates that CT2Rep accurately generates reports with content flow and medical terminology that closely resemble those written by radiologists, markedly surpassing the baseline established with CT-Net.


\subsection{Ablation Study on Longitudinal Data Utilization} \label{long}
We evaluated CT2RepLong's performance and the impact of incorporating prior data through an ablation study. Initially, we established a baseline by training CT2Rep solely on the longitudinal dataset without using any prior data. We then augmented this baseline with three strategies: utilizing embeddings from previous reports, previous volume embeddings, and their combination via simple fusion (excluding our longitudinal cross-attention mechanism). Table \ref{table} demonstrates the advantages of prior multimodal data and our unique cross-attention mechanism. The exception with R\(_{\textbf{L}}\) can be attributed to its emphasis on sequence length over the enriched content and diversity from longitudinal data integration. Besides, despite the limited size of longitudinal data—only 13\% of patients (see Sec. \ref{data})—CT2RepLong's performance was comparable with the original CT2Rep, illustrating its effectiveness, even with a constrained dataset.

\begin{figure}
    \centering
    \includegraphics[width=1\linewidth]{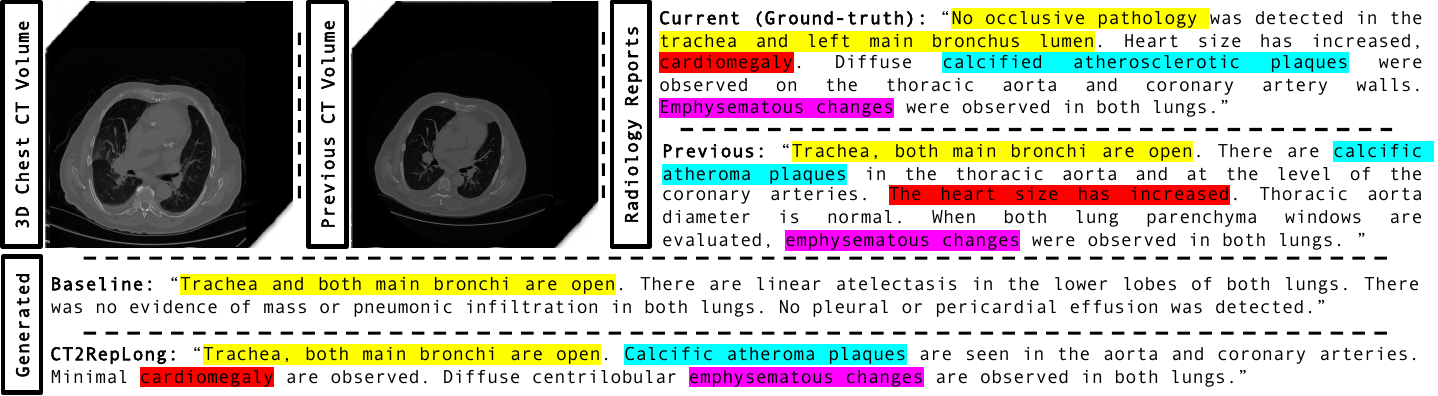}
    \caption{CT2RepLong surpasses the baseline, leveraging longitudinal data for enhanced medical detail accuracy, with related terms color-coded for clarity.}
    \label{ct2replongcase}

\end{figure}

\noindent\textbf{Case study.} A qualitative analysis of a random test case (Fig. \ref{ct2replongcase}) reveals that CT2RepLong significantly benefits from integrating longitudinal data. Key terms like “cardiomegaly” and “calcified atherosclerotic plaques” appeared in both current and previous reports, enriching the accuracy of the generated reports. Notably, terms missed by the baseline, such as “cardiomegaly” were included by CT2RepLong, aligning with the ground truth and demonstrating the enhanced reliability of report generation with our extension for utilizing longitudinal data.

\subsection{Implementation Details}
CT2Rep and the baseline method (see Sec. \ref{base}) were trained on 49,138 3D CT volumes and their corresponding reports (Sec. \ref{data}). We used the Adam optimizer with $\beta_1$ and $\beta_2$ hyperparameters set to 0.9 and 0.99, respectively. The learning rate was established at 0.00005 for the visual extractor and 0.0001 for the other parameters. A StepLR scheduler with a gamma of 0.1, a batch size of 1, and a maximum token count of 300 for the scheduler were employed. CT2RepLong and the ablation methods (Sec. \ref{long}) were trained on 28,441 pairs (Sec. \ref{data}), utilizing the same hyperparameters as CT2Rep. The training duration for all models was one week on a single NVIDIA A100 GPU, achieving 20 epochs. Inference takes approximately 35 seconds for CT2Rep and 50 seconds for CT2RepLong.

\section{Discussion and Conclusion}
\label{section:discussion}
In conclusion, we introduce CT2Rep, the first framework for automating 3D medical imaging report generation, with a focus on chest CT volumes. Leveraging an innovative auto-regressive causal transformer architecture and integrating relational memory, CT2Rep enhances accuracy in report generation. As the first of its kind, we establish a benchmark using the state-of-the-art vision encoder in 3D chest CT volume interpretation to showcase CT2Rep's effectiveness. Additionally, we extend its capabilities with longitudinal data integration, resulting in CT2RepLong, further enhancing context and accuracy. We make our trained models and code fully open-source to lay a solid foundation for further research. 

\begin{credits}
\subsubsection{\ackname} We extend our gratitude to the Helmut Horten Foundation for their invaluable support of our research. Additionally, we would like to express our sincere appreciation to Istanbul Medipol University for providing the CT-RATE dataset.

\subsubsection{\discintname}
The authors have no competing interests to declare that are relevant to the content of this article.
\end{credits}

\bibliographystyle{splncs04}
\bibliography{bibliography}

\clearpage
\label{section:supplementary}
\section*{Supplementary Material}

\definecolor{mycolor}{rgb}{0.90, 0.90, 0.90} 
\begin{table}[ht]
  \centering
  \setlength{\tabcolsep}{3.0pt} 
  \renewcommand{\arraystretch}{1.3} 
  \rowcolors{5}{mycolor}{white} 
  \begin{tabular}{
    @{}
    l
    S[table-format=1.3]
    S[table-format=1.3]
    S[table-format=1.3]
    S[table-format=1.3]
    S[table-format=1.3]
    S[table-format=1.3]
    S[table-format=1.3]
    @{}
  }
    \toprule
    & \multicolumn{3}{c}{\textbf{Base w/ CT-Net}} & \multicolumn{3}{c}{\textbf{CT2Rep (Ours)}} & {} \\
    \cmidrule(lr){2-4} \cmidrule(lr){5-7} 
    \textbf{Abnormality} & {\textbf{P}} & {\textbf{R}} & {\textbf{F1}} & {\textbf{P}} & {\textbf{R}} & {\textbf{F1}} & {\textbf{Test Set}} \\
    \midrule
    Medical material & 0.117 & 0.400 & 0.180 & 0.736 & 0.752 & 0.308 & 0.114 \\
    Arterial wall calcification & 0.795 & 0.635 & 0.706 & 0.928 & 0.356 & 0.675 & 0.267 \\
    Cardiomegaly & 0.088 & 0.336 & 0.139 & 0.657 & 0.686 & 0.555 & 0.108 \\
    Pericardial effusion & 0.996 & 0.820 & 0.899 & 0.769 & 0.920 & 0.786 & 0.074 \\
    Coronary artery wall calcification & 0.173 & 0.163 & 0.168 & 0.405 & 0.705 & 0.223 & 0.244 \\
    Hiatal hernia & 0.110 & 0.666 & 0.188 & 0.998 & 0.214 & 0.890 & 0.134 \\
    Lymphadenopathy & 0.932 & 0.380 & 0.540 & 0.641 & 0.645 & 0.555 & 0.266 \\
    Emphysema & 0.660 & 0.870 & 0.750 & 0.436 & 0.864 & 0.310 & 0.195 \\
    Atelectasis & 0.052 & 0.122 & 0.073 & 0.998 & 0.638 & 0.778 & 0.232 \\
    Lung nodule & 0.713 & 0.381 & 0.496 & 0.818 & 0.502 & 0.596 & 0.425 \\
    Lung opacity & 0.994 & 0.588 & 0.741 & 0.579 & 0.234 & 0.320 & 0.374 \\
    Pulmonary fibrotic sequela & 0.948 & 0.465 & 0.624 & 0.569 & 0.669 & 0.443 & 0.267 \\
    Pleural effusion & 0.137 & 0.649 & 0.226 & 0.570 & 0.396 & 0.443 & 0.126 \\
    Mosaic attenuation pattern & 0.289 & 0.230 & 0.256 & 0.984 & 0.242 & 0.386 & 0.078 \\
    Peribronchial thickening & 0.194 & 0.749 & 0.308 & 0.888 & 0.660 & 0.624 & 0.098 \\
    Consolidation & 0.293 & 0.795 & 0.428 & 0.804 & 0.417 & 0.526 & 0.172 \\
    Bronchiectasis & 0.782 & 0.502 & 0.601 & 0.756 & 0.425 & 0.516 & 0.099 \\
    Interlobular septal thickening & 0.969 & 0.811 & 0.883 & 0.953 & 0.547 & 0.684 & 0.071 \\
    \midrule
    \textbf{Mean} & 0.513 & 0.531 & 0.456 & 0.749 & 0.548 & 0.534 & 0.186 \\
    \bottomrule
  \end{tabular}
  \caption{Abnormality-based clinical efficacy metrics, including precision (P), recall (R), and F1 score, are showcased for generated reports by the CT-Net-based baseline and our CT2Rep method. CT2Rep's superior performance underscores the benefits of utilizing our novel auto-regressive causal transformer for 3D feature extraction, coupled with relational memory and memory-driven conditional layer normalization, to generate clinically accurate reports for 3D chest CT volumes. Additionally, the ratios of abnormalities in the test set are provided.}

\end{table}

\begin{table}[t]
  \centering
  \setlength{\tabcolsep}{3.0pt} 
  \renewcommand{\arraystretch}{1.3} 
  \rowcolors{5}{mycolor}{white} 
  \begin{tabular}{
    @{}
    l
    S[table-format=1.3]
    S[table-format=1.3]
    S[table-format=1.3]
    S[table-format=1.3]
    S[table-format=1.3]
    S[table-format=1.3]
    S[table-format=1.3]
    @{}
  }
    \toprule
    & \multicolumn{3}{c}{\textbf{Baseline}} & \multicolumn{3}{c}{\textbf{CT2RepLong}} & {} \\
    \cmidrule(lr){2-4} \cmidrule(lr){5-7} 
    \textbf{Abnormality} & {\textbf{P}} & {\textbf{R}} & {\textbf{F1}} & {\textbf{P}} & {\textbf{R}} & {\textbf{F1}} & {\textbf{Test Set}} \\
    \midrule
    Medical material & 0.490 & 0.593 & 0.492 & 0.910 & 0.617 & 0.555 & 0.233 \\
    Arterial wall calcification & 0.390 & 0.417 & 0.389 & 0.613 & 0.476 & 0.472 & 0.330 \\
    Cardiomegaly & 0.806 & 0.120 & 0.478 & 0.993 & 0.690 & 0.886 & 0.153 \\
    Pericardial effusion & 0.851 & 0.718 & 0.448 & 0.711 & 0.467 & 0.553 & 0.145 \\
    Coronary artery wall calcification & 0.947 & 0.250 & 0.569 & 0.750 & 0.657 & 0.664 & 0.296 \\
    Hiatal hernia & 0.486 & 0.316 & 0.472 & 0.943 & 0.805 & 0.472 & 0.148 \\
    Lymphadenopathy & 0.683 & 0.414 & 0.498 & 0.926 & 0.857 & 0.435 & 0.333 \\
    Emphysema & 0.658 & 0.396 & 0.567 & 1.000 & 0.635 & 0.777 & 0.300 \\
    Atelectasis & 0.577 & 0.416 & 0.572 & 0.454 & 0.287 & 0.457 & 0.258 \\
    Lung nodule & 0.864 & 0.347 & 0.740 & 0.703 & 0.348 & 0.585 & 0.453 \\
    Lung opacity & 0.563 & 0.894 & 0.490 & 0.480 & 0.153 & 0.315 & 0.535 \\
    Pulmonary fibrotic sequela & 0.640 & 0.570 & 0.539 & 0.541 & 0.766 & 0.548 & 0.295 \\
    Pleural effusion & 0.502 & 0.667 & 0.509 & 0.698 & 0.618 & 0.391 & 0.286 \\
    Mosaic attenuation pattern & 0.655 & 0.375 & 0.527 & 0.729 & 0.501 & 0.574 & 0.024 \\
    Peribronchial thickening & 0.695 & 0.536 & 0.569 & 0.974 & 0.955 & 0.958 & 0.168 \\
    Consolidation & 0.445 & 0.569 & 0.439 & 0.413 & 0.181 & 0.304 & 0.378 \\
    Bronchiectasis & 0.986 & 0.211 & 0.562 & 0.514 & 0.713 & 0.440 & 0.165 \\
    Interlobular septal thickening & 0.740 & 0.570 & 0.605 & 0.737 & 0.098 & 0.266 &  0.117 \\
    \midrule
    \textbf{Mean} & 0.666 & 0.465 & 0.525 & 0.727 & 0.511 & 0.536 & 0.257 \\
    \bottomrule
  \end{tabular}
  \caption{Clinical efficacy metrics (precision, recall, and F1) for each abnormality are showcased for reports generated by the baseline and our enhanced CT2RepLong method. Augmenting CT2Rep, CT2RepLong integrates a cross-attention multi-modal fusion module and longitudinal memory, effectively leveraging historical reports and volumes from previous visits. This method's superior performance over the baseline underlines the benefits of employing longitudinal multimodal data in producing clinically precise radiology reports for 3D chest CT volumes. The distribution of abnormalities in the test set is also detailed.}
\end{table}

\end{document}